\documentclass{article}
\usepackage{spconf,amsmath,epsfig}
\usepackage{enumitem}
\usepackage{booktabs}
\usepackage{caption}

\let\OLDthebibliography\thebibliography
\renewcommand\thebibliography[1]{
  \OLDthebibliography{#1}
  \setlength{\parskip}{0pt}
  \setlength{\itemsep}{0pt plus 0.3ex}
}

\begin{document}\sloppy

\def\x{{\mathbf x}}
\def\L{{\cal L}}

\title{An Order-Complexity Model for Aesthetic Quality Assessment of Homophony Music Performance}
%
\name{Xin Jin$^{1, 2}$, Wu Zhou$^1$, Jinyu Wang$^1$, Duo Xu$^{3, 4}$$^*$, Yiqing Rong$^1$ and Jialin Sun$^1$}
\address{$^1$Beijing Electronic Science and Technology Institute, 100070, Beijing, China, many33@126.com\\
$^2$Beijing Institute for General Artificial Intelligence (BIGAI), 100085, Beijing, China \\
$^3$Tongji University, 200082, Shanghai, China
$^4$Tianjin Conservatory of Music, 300171, Tianjin, China 
}

\maketitle

\begin{abstract}
    Although computational aesthetics evaluation has made certain achievements in many fields, its research of music performance remains to be explored. At present, subjective evaluation is still a ultimate method of music aesthetics research, but it will consume a lot of human and material resources. In addition, the music performance generated by AI is still mechanical, monotonous and lacking in beauty. In order to guide the generation task of AI music performance, and to improve the performance effect of human performers, this paper uses Birkhoff's aesthetic measure to propose a method of objective measurement of beauty. The main contributions of this paper are as follows: Firstly, we put forward an objective aesthetic evaluation method to measure the music performance aesthetic; Secondly, we propose 10 basic music features and 4 aesthetic music features. Experiments show that our method performs well on performance assessment.
\end{abstract}
\section{Introduction}
    Computational aesthetics evaluation \cite{galanter2012computational} enables computers to make quantitative aesthetic judgments on work of arts, which usually include architecture, painting and music. It can be used to compare the aesthetic feeling of different works of art for aesthetic quality assessment \cite{zhu2020metaiqa}.

    Music aesthetics has always been a difficult problem, which is proposed and attempt to be solved by mathematicians (going back to Pythagoras) up to music as emotional expression \cite{kivy1989sound} or music as language \cite{lerdahl1996generative}. This study believes that it is necessary to establish an objective method to quantify aesthetic feeling of music, and the aesthetic feeling of music is not completely subjective.

    Nowadays, compared with human performers, performance generated by AI sounds monotonous and lacks spirituality. For now, the machine may only have learned the distribution of a music performance attributes like dynamic and tempo, which may be one of the reasons why the performance of the machine is very mechanical and clumsy.

    \begin{figure}[htbp]

\centering
  \includegraphics[width=0.5\textwidth]{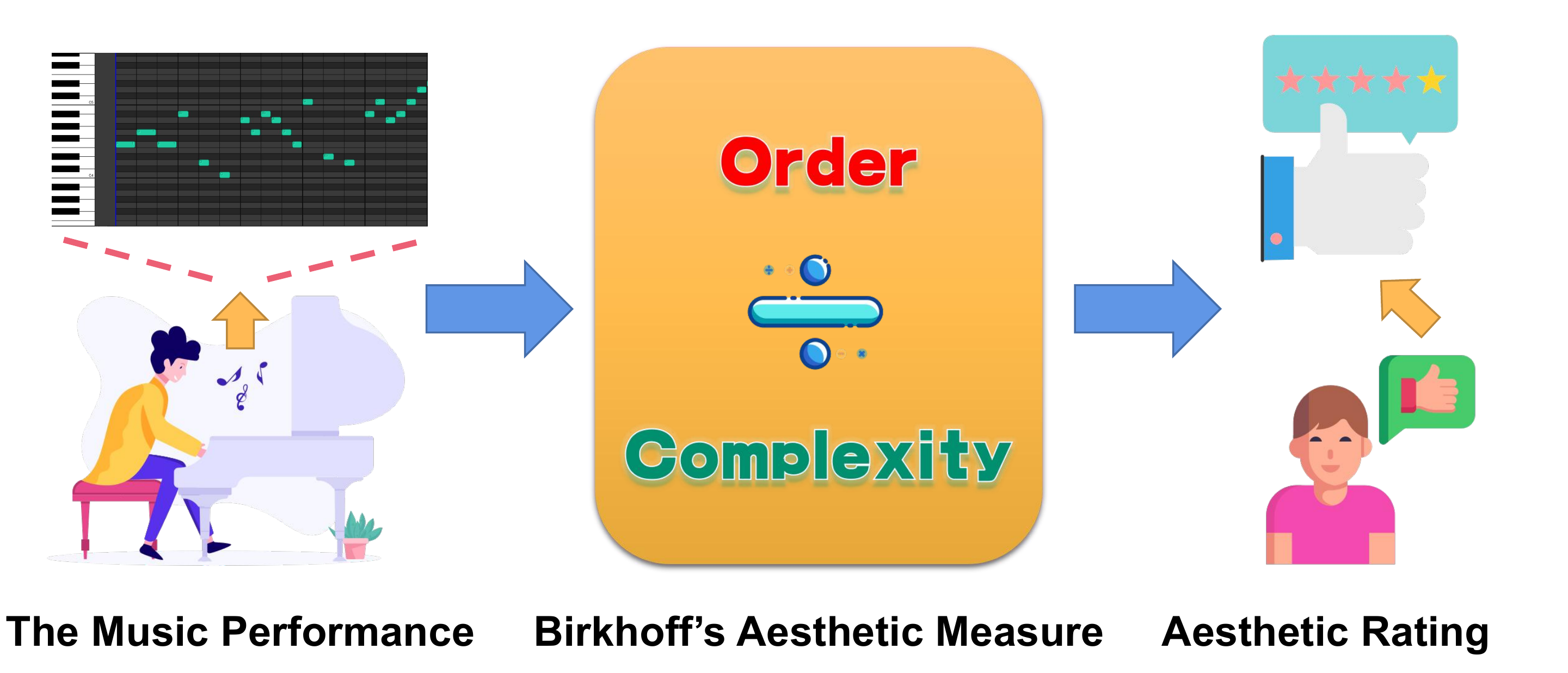}
  \caption{The quality of music performance can be evaluated through the Performance Aesthetic Assessment Model.}
  \vspace{-1.5em}
  \label{fig:introduction}
  
\end{figure}

    Current music performance data is not labeled with aesthetic ratings like AVA \cite{murray2012ava} in the field of image aesthetic evaluation. Due to the lack of datasets with subjective rating labels, we adopt objective traditional aesthetic measure.
    
    In this paper, Birkhoff’s method \cite{birkhoff2013aesthetic} was selected to conduct a study of aesthetic quality assessment. The reason why we choose the method is that it has a strong interpretability for beauty. Birkhoff formalizes the aesthetic measure of an object into the quotient between order and complexity:
    
    \begin{equation}
        M=\frac{O}{C} \label{1}
    \end{equation}

    Fig\ref{fig:introduction} shows the main tasks of this article. The main contributions of our work are as follows:

    \begin{itemize}[itemsep=2pt,topsep=0pt,parsep=0pt]
        \item  We put forward an aesthetic evaluation method to measure the music performance aesthetics.
        \item  We propose 10 basic music performance features and 4 aesthetic music features to facilitate the following music information retrieval research tasks.
    \end{itemize}

    \begin{figure*}[htbp]
    \centering
      \includegraphics[width=\textwidth]{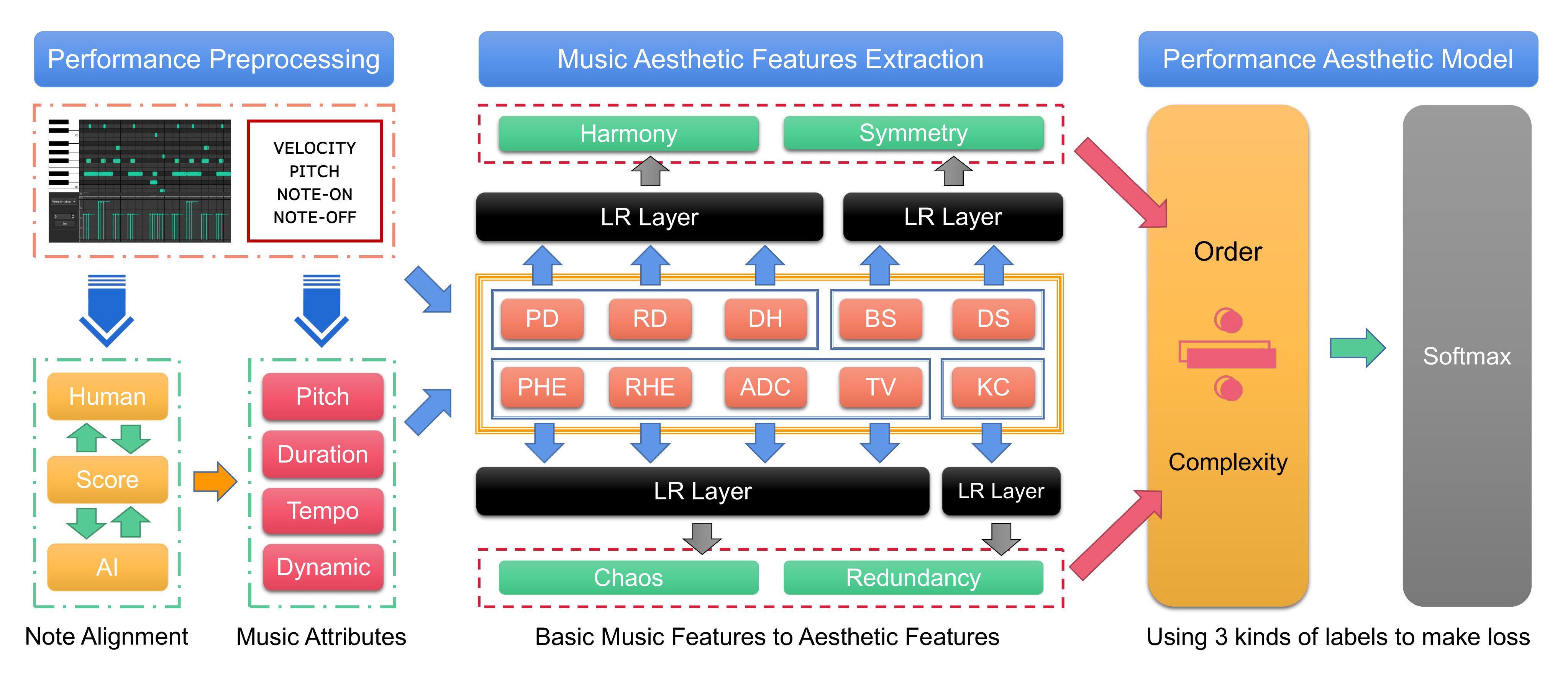}
      \caption{The overview of our approach. First of all, we align the notes of the score and performance midi. Then, we obtain music attributes from each note after the original performance and alignment. Next, we use the formula in section 3 to process music attributes and calculate ten basic music features. After that, we put 10 basic music features into the corresponding four logical regression layers (4 black blocks in the figure), and get 4 aesthetic features. Finally, we put four aesthetic features into our aesthetic assessment model, and then establish a loss through softmax to determine the weight.}
      \label{fig:network}
    \end{figure*}

\section{Related Work}
\subsection{Music Performance Generation}

    In this study, we focus on the tasks related to music performance generation. Maezawa et al.uses LSTM-based VAE \cite{maezawa2018deep} to generate expressive performance by giving conditional score, which renders performance with interpretation variations. VirtuosoNet \cite{jeong2019virtuosonet} is a relatively mature work. It has considered the important role of tempo, dynamic, and articulations in performance, which is considered to be a key issue. There are also some works such as DeepJ \cite{mao2018deepj} and MIDI-VAE \cite{brunner2018midi}. They also consider dynamic, but they still focus on the task of music score generation, so we will not elaborate here.

\subsection{Music Evaluation}
    
    The objective evaluation metrics are designed to unify the results of different models, so that the performance of different models can be compared. These metrics are usually statistics, and they mainly include 4 categories: pitch-related, rhythm-related, harmony-related and style transfer-related. These metric categories are usually used by AIGC models.
    
    Subjective evaluation methods mainly include listening test and visual analysis. The main methods of listening test include turning test \cite{hadjeres2017deepbach}, side by side rating \cite{haque2018conditional}, etc. These methods make a subjective evaluation of whether music is created by machines or people, as well as the aspects of music such as the harmony competition, the rhythm, the structure, the coherence and the overall rating.

    At present, there is a little work related to the evaluation of music aesthetics. Audio Oracle \cite{dubnov2011audio} (AO) uses Information Rate (IR) as an aesthetic measure. However, it cannot clarify what kind of specific aesthetic intention the system has. There is also a method to measure the beauty of music with Zipf's law \cite{manaris2002can}, which applies the rule of word frequency to music.

\section{our aesthetic assessment model}

\subsection{Formalization}

    Based on Birkhoff's measure, we propose four aesthetic features: harmony, symmetry, chaos and redundancy. We linearly combine the order measures of molecules and the complexity measures of denominators. Detailed measures explaination will be described in sections 3.2, 3.3, 3.4 and 3.5. Fig\ref{fig:network} shows the process of our work. The music aesthetic measure formula is as follows:

    \begin{equation}
        Aesthetic\ Measure=\frac{\omega_1H + \omega_2S + \theta_1}{\omega_3C + \omega_4R + \theta_2} \label{6}
    \end{equation}

    Where $H$ is harmony, $S$ is symmetry, $C$ is chaos and $R$ is redundancy. $\omega$ is the weight and $\theta$ is the constant.

\subsection{Harmony}

\subsubsection{Pitch Deviation \& Rhythm Deviation}

    Pitch deviation and rhythm deviation are used to judge the quality of the pitch and rhythm accuracy played by the performer. These two features compare the pitch attribute and rhythm attribute of each note in performance with the pitch attribute and rhythm attribute of each note in the original score, so as to calculate the pitch and rhythm deviation between the performer's performance and the original score. Performers must play the notes on the music correctly. The calculation formula of deviation is given below:

    \begin{equation}
        Deviation=\frac{\sum_{i=1}^n {\lvert X_i - T_i \rvert}}{\sum_{i=1}^n {\lvert T_i \rvert}}
    \end{equation}

    Where $i$ indicates the ith note in music, $n$ represents the total number of aligned notes, $X_i$ represents the attribute value of the ith note of perfomance and $T_i$ represents the attribute value of the ith note of score.
\subsubsection{Dynamic Harmony}

    According to GTTM theory \cite{lerdahl1996generative}, given a piece of music, the metric structure of the music can be obtained. The metrical structure is the rhythmic structure in a piece of music, which well describes the strong and weak structure in music.
    
    We use cosine similarity to measure the matching degree between dynamic and metrical structure value. The formula of dynamic harmony is as follows:
    
    \begin{equation}
    Dynamic \ Harmony = \frac{\sum_{i = 1}^{n} D_{i} \times M_{i}}{\sqrt{\sum_{i = 1}^{n}(D_{i})^{2} \times \sum_{i = 1}^{n}(M_{i})^{2}}}
    \end{equation}
    
    Where $i$ represents the ith note in the beginning of bar, $n$ represents the total number of aligned notes which in the beginning of bar, $D$ represents the vector of dynamic values and $M$ represents the vector of metrical structure values.

\subsection{Symmetry}

\subsubsection{Beat Skewness \& Dynamic Skewness}

    Skewness is used to measure the degree of asymmetry of distribution and it is the third-order normalized moment of the sample. In performance, the two more important factors are dynamic and beat. We calculate their skewness in symmetry and take the absolute value. The formula is given below:

    \begin{equation}
        Skewness=E\left[\left(\frac{X-\mu}{\sigma}\right)^{3}\right]
    \end{equation}

    Where $E$ represents mathematical expectation, $\mu$ represents the mean value, $\sigma$ representative standard deviation.

\subsection{Chaos}
     
\subsubsection{Pitch \& Rhythm Histogram Entropy}

    We use Shannon entropy and the definition of histogram entropy is given below:

    \begin{equation}
        Histogram \ Entropy=-\sum_{x\in\Omega}p(x)\log(x)
    \end{equation}

    It usually used as the measurement of the degree of chaos and uncertainty in the internal state of a system.

\subsubsection{Average Dynamic Changes}

    There are strong (f) and weak (p) symbols in the music score to express the dynamic when playing. The calculation of average dynamic changes is as follows:

    \begin{equation}
        Average \ Dynamic \ Changes=\frac{\sum_{i=2}^n {\lvert D_i - D_{i-1} \rvert}}{n-1}
    \end{equation}

     Where $i$ indicates the ith note in music, $n$ represents the total number of aligned notes. Starting from the second note, let each note make the dynamic difference between each note $D_i$ and the dynamic of the previous note $D_{i-1}$, add them up and divide by n-1.

\subsubsection{Tempo Variability}

    We design tempo variability to measure the chaos of tempo. Tempo variability is described as follows:

    \begin{equation}
        Tempo \ Variability=\sqrt{\frac{\sum_{i=1}^n (t_i - \bar{t})^2} {n-1}}
    \end{equation}

     Tempo variability measures standard deviation of the tempo in beats per minute. Where $t_i$ represents the ith tempo sample value $\bar{t}$ represents the mean of tempo.

\subsection{Redundancy}

\subsubsection{Kolmogrov Complexity}

    For a string $s$, Kolmogorov complexity $K(s)$ of the string $s$ refers to the shortest program to calculate the string $s$ on a computer. In essence, the Kolmogorov complexity of a string is the length of the final compressed version of the string.

     Lossless compression of music is a general measure of music's redundancy. We use lossless compression of music to describe it. It can be formalized as the following formula:
    
    \begin{equation}
        Kolmogorov\ Complexity=1 - \frac{K}{I_m}
    \end{equation}

    Where $K$ is the amount of information after lossless compression of music, and $I_m$ is the original amount of information of music.

\section{Implementation}
\subsection{Datasets}

    There are a few aligned datasets of music score and performance, we finally choose ASAP \cite{foscarin2020asap} as our dataset.

        \begin{table}[htbp]
        \centering
    \begin{tabular}{ccc}
    \toprule
    \textbf{Score} & \textbf{AI (VirtuosoNet)} & \textbf{Human} \\ \midrule
    234            & 234                       & 1060           \\ \bottomrule
    \end{tabular}

            \caption{The number of samples for each positive, negative and intermediate value.}
              \vspace{-1em}
            \label{tab:tab1}
    \end{table}

    As shown in Table\ref{tab:tab1}, we plan to select the score played by the machine in ASAP as the negative sample (not beautiful performance), and then select the performance of the corresponding score in ASAP as the positive sample (beautiful performance). Then, in order to obtain an intermediate value, we use VirtuosoNet's pre-trained model \cite{jeong2019virtuosonet}, take ASAP score as input, and render a corresponding number of performance as the intermediate value (just so so performance).

    In this way, we obtain three types of samples: the performance directly rendered by music scores, the AI generated the performance, and human performance.

    \subsection{Note Alignment}

     We need to use the attributes of each note in the score as reference values to ensure that the “improvisation" of performance is also based on evidence. To calculate features such as deviation, you need to align the notes of score and performance.

    We choose a score to performance algorithm proposed by Nakamura et al. \cite{nakamura2017performance}. The algorithm processes notes executed asynchronously, as well as missing and additional notes in performance, and returns a list of note-to-note matches.

    Although the note alignment algorithm is used, it shows that our alignment still has errors. So we adopt the refinement algorithm in VirtuosoNet \cite{jeong2019virtuosonet} to improve the note alignment. 

    \subsection{Basic Features Calculation}

     To calculate the basic music features, We must get the attributes of the note. We use music21\footnotemark{} and jSymbolic \cite{mckay2018jsymbolic} to get the pitch, duration, tempo and dynamic of the notes.
    
    We use absolute pitch for subsequent calculation. According to the formula in 3.2.1, we calculate the deviation on the aligned notes to get PD. Then we calculate the absolute pitch histogram and use the entropy formula to calculate the PHE. Since then, the pitch related features have been calculated.

    \begin{table*}[htbp]
\centering
\renewcommand\arraystretch{1.25}
\tabcolsep=0.22cm
\begin{tabular}{@{}c|cccccccccc|cccc@{}}
\toprule
\textbf{Dataset} & \textbf{PD$\downarrow$}   & \textbf{RD$\downarrow$}   & \textbf{DH$\uparrow$}   & \textbf{BS$\downarrow$}   & \textbf{DS$\downarrow$}   & \textbf{PHE$\downarrow$}  & \textbf{RHE$\downarrow$}  & \textbf{ADC$\uparrow$}  & \textbf{TV$\uparrow$}   & \textbf{KC$\downarrow$}  & \textbf{H$\uparrow$} & \textbf{S$\uparrow$} & \textbf{C$\uparrow$} & \textbf{R$\uparrow$} \\ \midrule
Score            & \textbf{0.09} & 0.32          & 0.19          & 0.76          & 0.65          & 3.30          & 1.62          & 0.87          & 3.30          & 0.77     & -0.77            & 2.71              & 9.40           & -0.48      \\
AI               & 0.12          & \textbf{0.22} & 0.56          & 0.59          & 0.41          & \textbf{3.29} & 1.44          & 5.25          & 3.20          & 0.75     & 0.56             & 0.91              & 5.15           & -0.46      \\
Human            & 0.13          & 0.29          & \textbf{0.78} & \textbf{0.33} & \textbf{0.20} & 3.63          & \textbf{1.01} & \textbf{10.4} & \textbf{7.26} & \textbf{0.72}  & \textbf{2.76}    & \textbf{3.27}     & \textbf{11.65} & \textbf{1.01}\\ \bottomrule
\end{tabular}
 \vspace{0.2cm}

\caption{The above is the experimental data on three datasets, which are displayed by the mean value of ten basic music features and four aesthetic features. \textbf{The bold data in the table has higher aesthetic significance.} It is obvious from the table that human's performance has better aesthetic significance than the music score and the performance generated by AI.}
\label{tab:tab3}
\end{table*}

    For RD, we can directly use music21 to obtain the duration attribute of the note and calculate the rhythm deviation directly. For the calculation method of RHE, we use a coding form of rhythm in jSymbolic, which encodes rhythm into 12 different lengths to eliminate the difference between performance and score to obtain a normalized rhythm histogram.
    
\begin{figure}[htbp]
      \centering
      \includegraphics[width=0.5\textwidth]{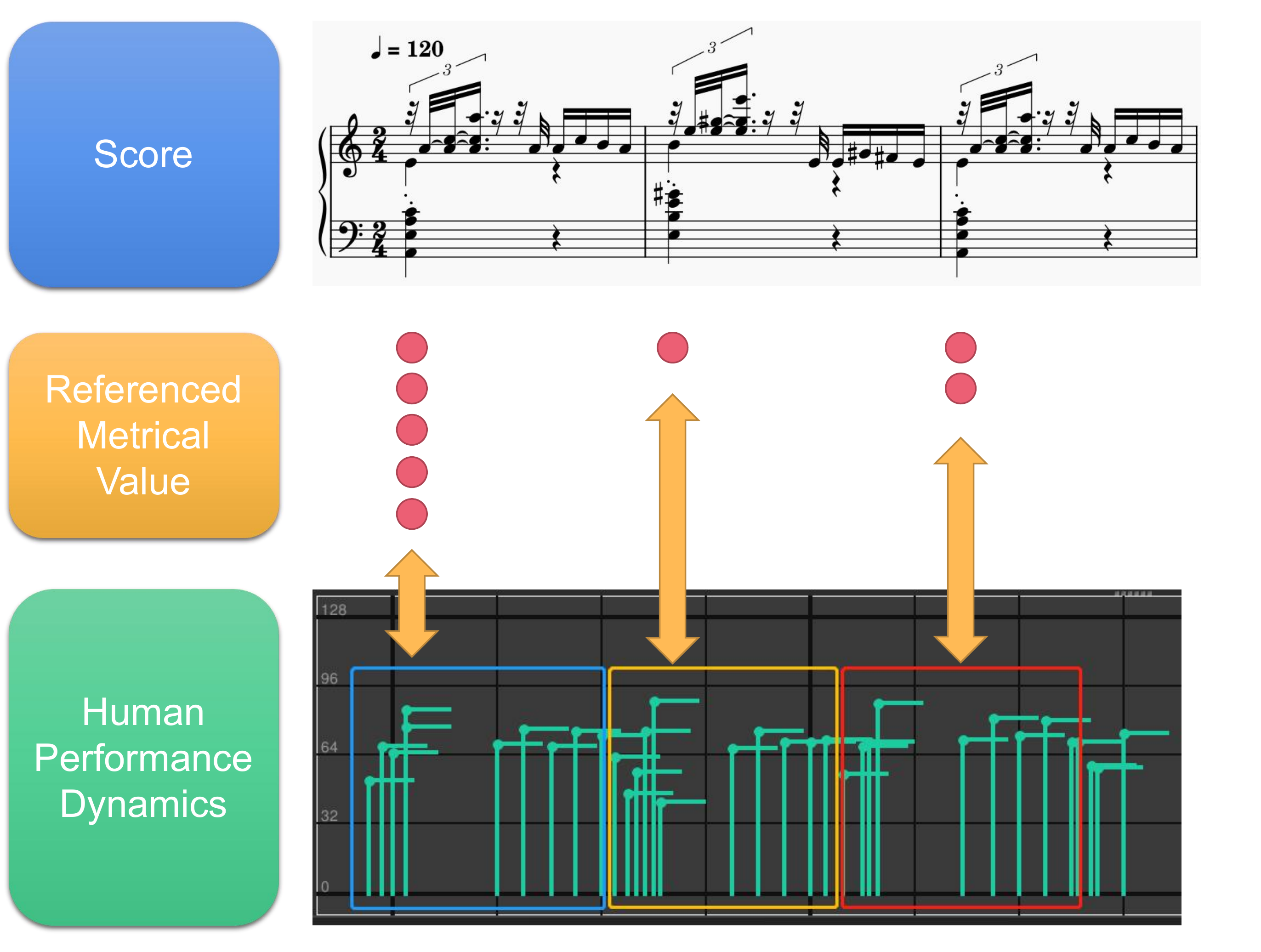}
      \caption{Calculating the mean value of dynamic in each bar to match the referenced metrical value.}
      \label{fig:matching}
    \end{figure}

    We divide velocity (dynamic) into five equal parts, setting 0-24 as level 0, 25-50 as level 1, 51-76 as level 2, 76-101 as level 3, and 102-127 as level 4. Since the reference dynamic value is labeled by bar, we take the mean of the dynamic values of all notes in each bar to form an n-dimensional vector, where n is the number of bars. The referenced dynamic is also a n-dimensional vector. Then we calculate the cosine similarity between the referenced dynamic value and the dynamic in performance to obtain the dynamic harmony (DH). Similarly, we can calculate the values of DS and ADC according to the formula in previous.  Fig\ref{fig:matching} briefly shows our approach.

    \vspace{-0.2cm}
    
    \vspace{-0.5em}
    For tempo, we use the concept of beat in jSymbolic to describe the tempo of music, which is essentially another way to describe the rhythm. We use the beat histogram to calculate BS. For the feature of tempo variation, we directly choose the implementation of jSymbolic to get the value of TV.
    \footnotetext{https://web.mit.edu/music21/}

    Finally, only KC has not been calculated. We use musescore3 to render the midi files, render them into wav audio files, and sample at the frequency of 44100Hz to ensure that the music information is lossless. Then, we refer to Monkey's Audio's\footnotemark{}\footnotetext{https://www.monkeysaudio.com/} lossless compression method to compress music into ape format, so as to calculate KC.
    
     \begin{figure*}[htbp]

      \centering
      \includegraphics[width=\textwidth]{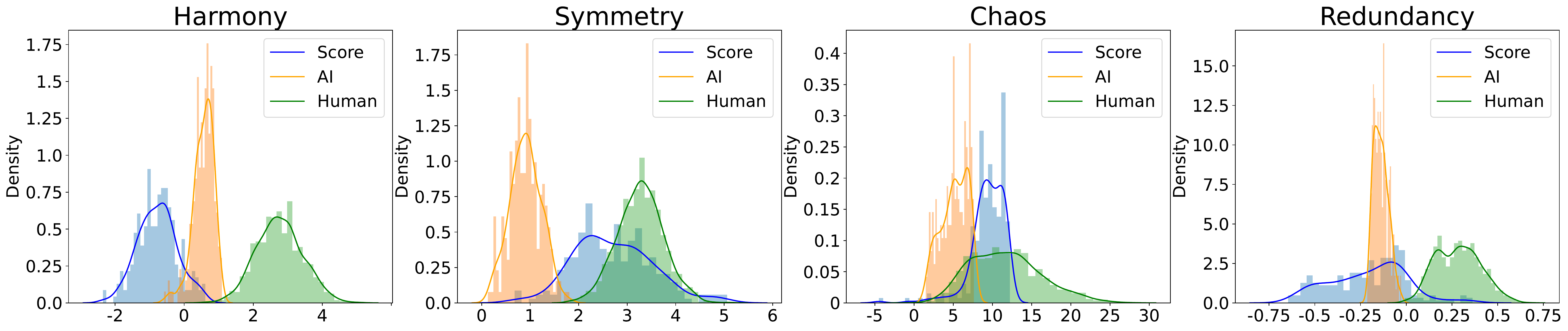}
      \caption{The distribution of 4 aesthetic features, which are obtained through logistic regression. Among them, Harmony is better differentiated, while Symmetry, Chaos, Redundancy are worse differentiated, showing the necessity of using O/C model.}
      \label{fig:AFD}
    \end{figure*}

    \subsection{The Aesthetic Features \& Loss Optimization}

     We combine our basic music features in a logistic regression model to get four aesthetic features. We normalize the value of basic music features, and then put all samples into the logistic regression model for training.

     Then we put the original value of the aesthetic features into the formula in 3.1, and then calculate its softmax. We choose the cross entropy loss function, and use the gradient descent method to minimize the loss function. By setting the learning rate to 0.01, after 1000 iterations, the loss function converges and the model parameters are solved.
     
\section{experiments}
\subsection{Main Results \& Ablation Study}

    The main results are divided into two parts. One is to introduce feature distribution, and the other is to use the metric in section 5.1 to evaluate the performance of the model.

    In Table\ref{tab:tab3}, we have calculated the mean value of 10 basic music features and 4 aesthetic features on three data sets, which makes “beauty" interpretable. Let's analyze each basic feature in detail: 
    
    Score has the lowest PD, while human's PD is the highest. According to common sense, human plays “not very accurately", but this just shows that score's playing is very mechanical, and human's playing can be more flexible when it has certain playing skills. So is RD. DH is the focus of our research. The high DH value of human shows the importance of dynamic performance to conform to the metrical structure. The value of BS and DS in human playing is significantly lower, which indicates that the beat and dynamic played by human are more symmetrical and aesthetic than AI and score. Humans’ performance higher PHE and lower RHE, indicating that people may prefer to add some of their own playing skills, such as decorative tones. High ADC and TV values also indicate the diversity of human playing, and low KC values also indicate the low redundancy of human playing.

    \begin{figure}[htbp]
      \centering
      \includegraphics[width=0.35\textwidth]{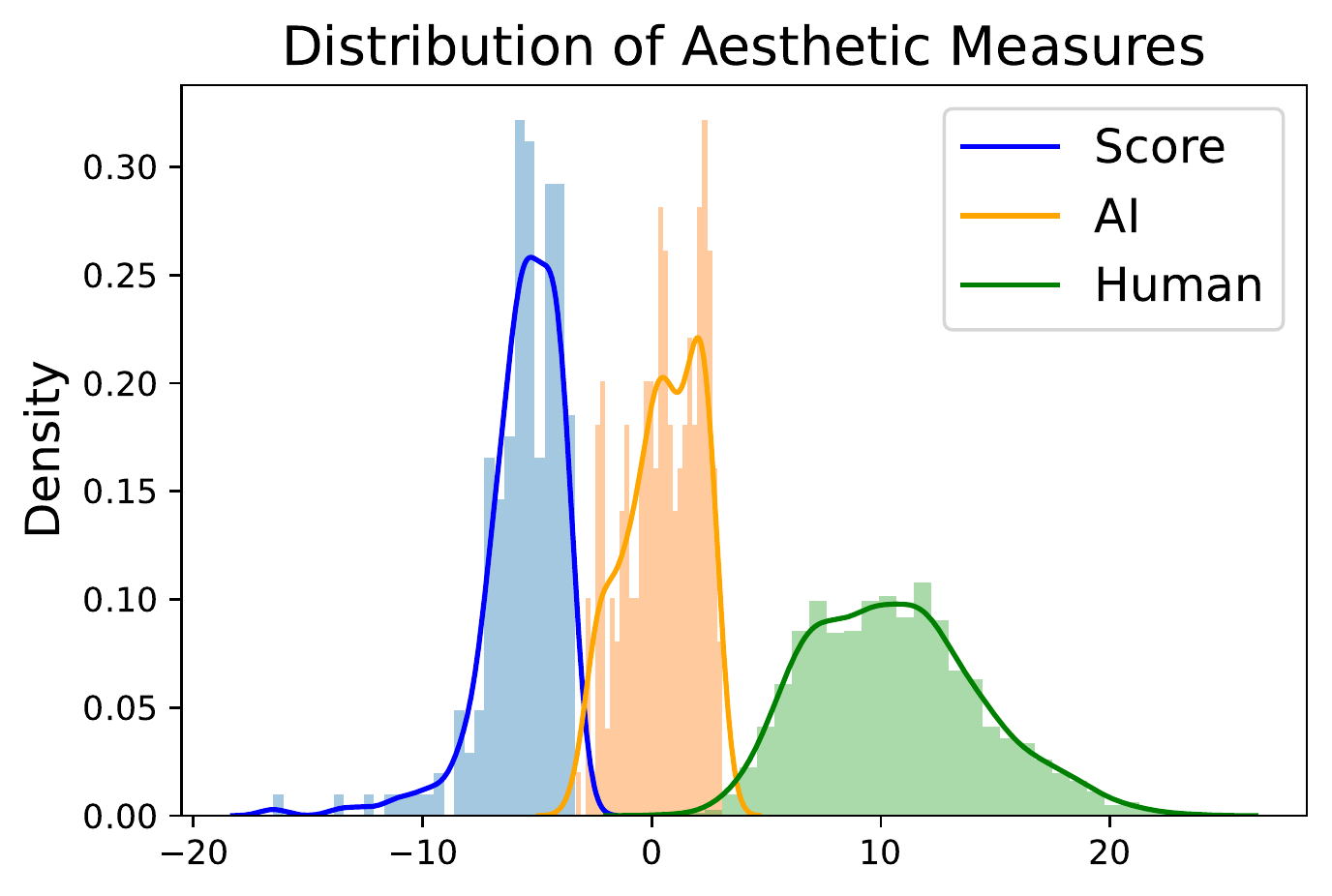}
      \caption{The 3 distribution of Birkhoff's aesthetic measure.}
      \vspace{-0.3cm}
      \label{fig:distribution}
    \end{figure}

\begin{table*}[htbp]
\centering
\tabcolsep=0.64cm
\begin{tabular}{@{}c|ccccc@{}}
\toprule
\textbf{metric}  & \textbf{w/o harmony} & \textbf{w/o symmetry} & \textbf{w/o chaos} & \textbf{w/o redundancy} & \textbf{our full model}\\ \midrule
accuracy             & 77.6\%               & 88.7\%                & 82.4\%             & 86.5\%        & \textbf{92.3\%}          \\
kappa                 & 0.579                & 0.791                 & 0.653              & 0.726         & \textbf{0.874}          \\ \bottomrule
\end{tabular}
\vspace{0.1cm}
\caption{The value of kappa coefficient ranges from -1 to 1. The larger the kappa coefficient, the better the performance of the model. The results show that our full model has the best performance.}

\label{tab:tab4}
\end{table*}

    Fig\ref{fig:AFD} shows the distribution of four aesthetic features. Unlike the direct use of deep learning and the use of neural networks to classify, the values generated in each step can be consulted to explain where “beauty" comes from. Fig\ref{fig:distribution} shows the distribution of aesthetic scores on the three datasets. As shown in the figure, the intersection of the distribution is small, which proves that our model performances very well.

    We also conducts ablation experiments to remove harmony, symmetry, chaos and redundancy respectively to train four different models. We compare them with the full model. We choose accuracy and kappa as the evaluation metrics. The results are shown in Table\ref{tab:tab4}.

\subsection{Subjective Evaluation}

    Subjective evaluation is still the standard to test the effectiveness of aesthetic scoring. We randomly selected 15 (5 * 3) pieces of music from score, AI and human datasets, and each piece lasts about 15 seconds. Volunteers participating in the subjective experiment need to choose one in each group (5 groups) that they think is the best.

    We find a total of 31 volunteers with 155 samples, and each volunteer listens to all the pieces. Among them, 32 samples (20.6\%) prefer score performance, 39 samples (25.2\%) prefer AI performance, and 84 samples (54.2\%) prefer human performance.

\section{conclusion}

    In summary, we propose an aesthetic model based on Birkhoff's order-complexity measure to assess the beauty of homophony music performance. For the AI music generation task, we propose 10 basic music features and 4 aesthetic features to evaluate the beauty of the generated music, which will help improve the quality of the AI music generation task. However, we have to admit that our model also has defects, such as the concept of symmetry, and our model does not consider “creativity". Moreover, the aesthetic study of music audio quality assessment is worth exploring in future.

\section*{Acknowledgment}
    We thank the ACs, reviewers and the AI Music Team of BIGAI. This work is partially supported by the Natural Science Foundation of China (62072014\& 62106118), the Project of Philosophy and Social Science Research, Ministry of Education of China (20YJC760115), the Open Fund Project of the State Key Laboratory of Complex System Management and Control (2022111), 

\bibliographystyle{IEEEbib}
\bibliography{icme2023template}

\end{document}